\documentclass[runningheads]{llncs}

\usepackage[utf8]{inputenc}
\usepackage{graphicx}
\graphicspath{{images/}}

\usepackage{hyperref}
\usepackage{cite}
\usepackage{soul}
\usepackage{url}
\usepackage{amsmath}
\usepackage{booktabs}
\usepackage{xcolor}
\definecolor{sodablue}{rgb}{0.05, 0.05, 0.75}
\definecolor{scalablue}{rgb}{0.20, 0.00, 0.65}
\definecolor{leanblue}{rgb}{0.20, 0.00, 0.65}
\definecolor{affirmativegreen}{rgb}{0, 0.5, 0}
\definecolor{negativered}{rgb}{0.5, 0, 0}

\usepackage{xspace}
\usepackage{mathtools}
\usepackage{orcidlink}

\usepackage{listings}
\lstdefinelanguage{Soda}{
    morekeywords={lambda, any, fun, def, if, then, else, match, case, class, extends, abstract, end, datatype, data, inductive, this, subtype, supertype, false, true, not, and, or, package, import, theorem, directive, @tailrec, @override, @new},
    sensitive=true,
    morecomment=[s]{/*}{*/},
    morestring=[b]"
}
\newcommand{\srccode}[1]{\texttt{{#1}}}
\newcommand{\basicType}[1]{{\color{black}\srccode{#1}}\xspace}
\newcommand{\annotation}[1]{{\color{brown}\srccode{#1}}\xspace}

\newcommand{\reservedWordSoda}[1]{{\color{sodablue}\srccode{#1}}\xspace}

\newcommand{\sodatypearrow}{\ensuremath{\rightarrow}\xspace}
\newcommand{\sodalambdaarrow}{\ensuremath{\longrightarrow}\xspace}
\newcommand{\sodacasearrow}{\ensuremath{\Longrightarrow}\xspace}

\newcommand{\sodalambda}{\reservedWordSoda{lambda}}

\newcommand{\sodaif}{\reservedWordSoda{if}}
\newcommand{\sodathen}{\reservedWordSoda{then}}
\newcommand{\sodaelse}{\reservedWordSoda{else}}

\newcommand{\sodamatch}{\reservedWordSoda{match}}
\newcommand{\sodacase}{\reservedWordSoda{case}}

\newcommand{\sodaclass}{\reservedWordSoda{class}}
\newcommand{\sodaextends}{\reservedWordSoda{extends}}
\newcommand{\sodaabstract}{\reservedWordSoda{abstract}}
\newcommand{\sodaendclass}{\reservedWordSoda{end}}
\newcommand{\sodadatatype}{\reservedWordSoda{datatype}}
\newcommand{\sodathis}{\reservedWordSoda{this}}
\newcommand{\sodasubtype}{\reservedWordSoda{subtype}}
\newcommand{\sodasupertype}{\reservedWordSoda{supertype}}

\newcommand{\sodanot}{\reservedWordSoda{not}}
\newcommand{\sodaand}{\reservedWordSoda{and}}
\newcommand{\sodaor}{\reservedWordSoda{or}}

\newcommand{\sodapackage}{\reservedWordSoda{package}}
\newcommand{\sodaimport}{\reservedWordSoda{import}}

\newcommand{\sodadirective}{\reservedWordSoda{directive}}

\newcommand{\sodatailrec}{\annotation{@tailrec}}

\newcommand{\sodalessthancolon}{\reservedWordSoda{<:}}
\newcommand{\sodagreaterthancolon}{\reservedWordSoda{>:}}

\newcommand{\reservedWordScala}[1]{{\color{scalablue}\srccode{#1}}\xspace}
\newcommand{\scalaType}[1]{\basicType{{#1}}}

\newcommand{\scalacase}{\reservedWordScala{case}}
\newcommand{\scalaclass}{\reservedWordScala{class}}
\newcommand{\scaladef}{\reservedWordScala{def}}
\newcommand{\scalalazy}{\reservedWordScala{lazy}}
\newcommand{\scalasealed}{\reservedWordScala{sealed}}

\newcommand{\scalatrait}{\reservedWordScala{trait}}
\newcommand{\scalaobject}{\reservedWordScala{object}}
\newcommand{\scalaval}{\reservedWordScala{val}}

\newcommand{\reservedWordLean}[1]{{\color{leanblue}\srccode{#1}}\xspace}
\newcommand{\leanType}[1]{\basicType{{#1}}}

\newcommand{\leandef}{\reservedWordLean{def}}

\newcommand{\leanmatch}{\reservedWordLean{match}}
\newcommand{\leanwith}{\reservedWordLean{with}}
\newcommand{\leancasebar}{\reservedWordLean{|}}
\newcommand{\leancasearrow}{\ensuremath{\Rightarrow}\xspace}
\newcommand{\leaninductive}{\reservedWordLean{inductive}}
\newcommand{\leanclass}{\reservedWordLean{class}}
\newcommand{\leannamespace}{\reservedWordLean{namespace}}
\newcommand{\leanwhere}{\reservedWordLean{where}}
\newcommand{\leanderiving}{\reservedWordLean{deriving}}

\newcommand{\newsoda}[1]{`{#1}'}

\newcommand{\Soda}{\textsc{Soda}\xspace}

\newcommand{\Yes}{{\color{affirmativegreen} Yes}}
\newcommand{\No}{{\color{negativered} No}}

\newcommand{\orcidIDlink}[1]{\href{https://orcid.org/#1}{\orcidlink{#1} \texttt{{https://orcid.org/#1}}}}

\begin{document}

    \title{\Soda: An Object-Oriented Functional Language for Specifying Human-Centered Problems}

    \author{Julian~Alfredo~Mendez\thanks{~\orcidIDlink{0000-0002-7383-0529}}}

    \authorrunning{J. A. Mendez}
    \titlerunning{An Object-Oriented Functional Language for Specifying Human-Centered Problems}

    \institute{Ume{\aa} University, Sweden\\
    \email{julian.mendez@cs.umu.se}
    }

    \maketitle

    \begin{abstract}
        We present \Soda (Symbolic Objective Descriptive Analysis), a language that helps to treat qualities and quantities in a natural way and greatly simplifies the task of checking correctness.
        We present key properties for the language motivated by the design of a descriptive language to encode complex requirements on computer systems, and we explain how these key properties must be addressed to model these requirements with simple definitions.
        We give an overview of a tool that helps to describe problems in an easy way, which we consider more transparent and less error-prone.
    \end{abstract}



    \section{Introduction}

    Understanding how artificial intelligence (AI) agents work can be challenging because AI algorithms are complex and their reasoning is opaque.
    Although transparency is often seen as a requirement, realistically, it might not always be possible due to privacy or security concerns. However, the need to ensure that a system operates within moral bounds remains.
    At the same time, validation and verification procedures highly depend on the specific contextual interpretations that have been employed to ground abstract principles (e.g., fairness or privacy) into the concrete functionalities of an agent~\cite{AlerTubella-2019}.

    Verification can be a difficult or unfeasible task, and even when achieved, its specifications can be difficult to understand.
    Verification is only as strong as the assumptions that underlie the specification, which means that specifying assumptions and analyzing these specifications is crucial for verification~\cite{Dennis-2016}.
    Given that AI mostly operates in environments that are at best partially known by the system designer, and that properties are often discussed at a high level of abstraction by stakeholders without a background in formal languages, specification languages need to be easily understood by different stakeholders.

    From a technical point of view, unit tests are crucial in ensuring the quality of a piece of software~\cite{Zhu-1997}.
    Test-driven development (TDD), which is a software development process in which developers create test cases together with the main development, has been shown to be more productive than other more conventional development techniques~\cite{Erdogmus-2005}.
    We propose going one step further by introducing theorems together with the code, which in turn becomes more reliable.

    Our contribution is to present \Soda~\footnote{\url{https://julianmendez.github.io/soda}} (Symbolic Objective Descriptive Analysis), a new descriptive language based on widely adopted concepts such as modelling with classes and functional definitions.
    In this context, by descriptive language, we mean a language based on descriptions rather than processes, which seems closer to a specification language than an implementation language.
    The set of basic constructs suffices to model complex requirements, but is small enough to be immediately understood by a larger group of stakeholders.

    In this paper, we address the following research question, which is composed of two parts:

    \textbf{RQ}: Can we design a \textbf{descriptive language} to encode requirements for AI systems such that:
    \begin{itemize}
        \item \textbf{RQ.1:} the language or a fragment is \textbf{formally verifiable}, and
        \item \textbf{RQ.2:} it is easily integrated into \textbf{state-of-the-art technology}?
    \end{itemize}

    In the following sections, we show our approach to this question.

    \section{Language Description}

    In this section, we present key properties for the language motivated by \textbf{RQ}, and we explain how these key properties must be addressed to model requirements with simple definitions.

    \subsection{Key Properties}

    One of the key properties of the language is that it should be used to formalize requirements in an intuitive way, and one of the most effective ways to do this is to use types.
    We want the language to be \emph{statically typed}, because static typing simplifies type checking, which in turn helps prevent formalization errors.
    To avoid errors caused by side effects~\cite{Martin-2009},
    we make variables \emph{immutable}, which is closer to their standard mathematical interpretation.

    We want to relate qualities and quantities to describe complex conditions.
    We expect to use the same language to model a problem containing hierarchies, like a resource access monitoring agent,
    and a problem containing measures, like a price monitoring agent.
    To do this, the language has to be \emph{expressive} and \emph{general}.

    Standard computer languages include a considerable number of reserved words and basic types, which sometimes hinder understanding.
    More reserved words usually bring more combinations and nuances to their use, and to simplify their comprehension, it is convenient for the language to have a \emph{small} set of constructs.

    Alongside the aforementioned properties, we want the language to be used to evaluate effectively whether properties hold.
    For that, the language should be \emph{easy to prototype}, and its prototypes should be \emph{human-level efficient}, which means efficient enough for its expected use.

    \subsection{Main Constructs}

    The main constructs are presented to ensure the requirements.
    We have chosen constructs that look similar to those used in popular programming languages.

    Since \Soda is \emph{statically typed}, it needs a type definition construct.
    Let us name it \newsoda{:} (colon).
    The syntax is $x : A$, meaning that $x$ is of type $A$.

    Due to \emph{immutability}, we want to define constants and functions, but not variables in the computer science sense.
    In the following, we mention variables in the mathematical sense when referring to lambda expressions.
    To define a constant or a function, we need a construct.
    Let us name it \newsoda{=} (equals sign).
    The notation is
    \begin{center}
        $f (x_{1} : A_{1}) \ldots (x_{n} : A_{n}) : A = e$
    \end{center}
    where $f$ is the \emph{function name}, each $x_{i}$ $(1 \leq i \leq n)$ is a parameter of type $A_{i}$, and $e$ is an expression of type $A$.
    A function $f$ without parameters is called a \emph{constant}.
    A function can be called using named parameters with the \newsoda{:=} symbol.
    For example, $f (x : Int) (y : Int) : Int$ can be invoked as $f (x := 0) (y := 1)$.

    Most current programming languages include lambda expressions, which are anonymous functions based on lambda calculus~\cite{Church-1936}.
    We have included lambda expressions in \Soda because they have been highly adopted.
    The notation
    \begin{center}
        $\sodalambda \ x \sodalambdaarrow f(x)$
    \end{center}
    corresponds to $(\lambda x).f(x)$ or $\lambda x \to f(x)$ in the literature.
    Since we work with typed lambda calculus, the type needs to be specified when it cannot be inferred, and we denote it by $\sodalambda \ (x : A) \sodalambdaarrow f(x)$.

    Since the language is \emph{expressive} and \emph{general}, it includes standard operations from mathematics, such as \newsoda{+}, \newsoda{-}, \newsoda{*}, \newsoda{/}, for arithmetic, and \newsoda{\sodanot}, \newsoda{\sodaand}, \newsoda{\sodaor} for logic, with the usual meaning.
    Logic functions are evaluated with lazy evaluation.
    Therefore, when $a$ \sodaand $b$ is evaluated, $a$ is evaluated first, and if $a$ is already false, $b$ is not evaluated.
    Analogously for \sodaor, if the first value is already true.
    Lazy evaluation can also be used to compute functions that would otherwise be undefined, and this is done by using defined parts on the left to prevent undefined parts on the right.
    If the computations have no side effects, the result of computing with or without lazy evaluation is exactly the same, but the time needed is not.

    Note that the language can define recursive functions over finite structures, as mainstream purely functional programming languages do.
    This gives an expressive power that is enough to model human-understandable constraints.

    To define piecewise functions, we have \newsoda{\sodaif-\sodathen-\sodaelse} structures, with the notation
    \begin{center}
        \sodaif $b$ \ \ \sodathen $e_{1}$ \ \ \sodaelse $e_{2}$
    \end{center}
    where $b$ is a Boolean expression, and $e_{1}$ and $e_{2}$ are expressions of the same type.
    The interpretation is standard, and the result is $e_{1}$ if $b$ is true, and $e_{2}$ otherwise.

    The pattern matching construct, called \newsoda{\sodamatch-\sodacase}, has the format:
    \begin{center}
        \sodamatch $x$ \ \ \sodacase $p_{1}$ \sodacasearrow $e_{1}$ \ \ \ldots \ \ \sodacase $p_{n}$ \sodacasearrow $e_{n}$
    \end{center}
    where $x$ is a variable to match, $p_{i}$ are patterns, and $e_{i}$ are expressions for $1 \leq i \leq n$.
    The type of the match structure is the most specific supertype of the $e_{i}$ expressions.
    The $p_{i}$ patterns could be of different types, but they should be constructors that possibly contain construction variables.
    In fact, we use pattern matching to use extractors for object deconstruction, which can also be used for type checking.
    Although the notation \sodaif-\sodathen-\sodaelse could be defined as a pattern matching structure, we decided to keep it because it is more concise and more universally recognizable.

    We find it relevant to highlight that the constructs we present in this section are meant to create small functions, which improve readability~\cite{Martin-2009}. They also require the use of function names in intermediate computations to create accurate documentation for specifications.

    \subsection{Types and Classes}

    Humans classify concepts into categories using features that help describe and reason.
    In \Soda, we use types and classes to model objects that have attributes.
    They help us model ethical values like privacy and fairness, especially in relation to regulations.

    There are some differences in view of whether it is convenient to have classes instantiated by objects (like in Java~\cite{Java-1995}), or whether it is better to have modules, where functions are imported (like in Haskell~\cite{Haskell-1990}).
    We compromise between these two options because the objects created with the classes of \Soda are immutable, and they work as namespaces for modules or to specify how to retrieve attributes in objects.

    We distinguish between a type and a class, as is usual in the literature~\cite{Atkinson-1990}: an object has a \emph{type}, and the type describes what the object can do, but not how. In contrast, a \emph{class} provides the implementation for an object.

    \Soda allows for the definition of algebraic data types, particularly inductive data types.
    The reserved word for this definition is \newsoda{\sodadatatype}, followed by the name of the data type.
    The constructors are declared on the subsequent lines.
    Each constructor takes zero or more parameters and returns an instance of the type being defined.
    Every parameter requires both a name and a type, with the parameters and the final return type separated by the type arrow (\sodatypearrow).
    This structure is denoted as follows:

    \begin{center}
        \begin{tabular}{l}
            \sodadatatype \(A\)                                                                                          \\
            \ \ $C_{1}$ : $(p_{1,1} : T_{1,1}) \rightarrow \ldots \rightarrow (p_{1,k_{1}} : T_{1,k_{1}}) \rightarrow A$ \\
            \ \ $\dots $                                                                                                 \\
            \ \ $C_{m}$ : $(p_{m,1} : T_{m,1}) \rightarrow \ldots \rightarrow (p_{m,k_{m}} : T_{m,k_{m}}) \rightarrow A$ \\
        \end{tabular}
    \end{center}
    where $m$ is the number of constructors. For each $1 \leq i \leq m$, $C_{i}$ represents a constructor and $k_{i}$ is a non-negative integer representing its number of parameters. When $k_{i} > 0$, then for each $1 \leq j \leq k_{i}$, $p_{i,j}$ and $T_{i,j}$ denote the parameter name and its respective type.

    For example:

    \begin{lstlisting}[label={lst:example-of-datatype}]
datatype NaturalNumber
  Zero : NaturalNumber
  Successor : (n : NaturalNumber) -> NaturalNumber
    \end{lstlisting}

    As most designers and programmers are familiar with object-oriented programming, we choose to build classes with a construct called \newsoda{\sodaclass}.
    We adhere to the Open-Closed Principle, where ``software entities (classes, modules, functions, etc.) should be open for extension, but closed for modification''~\cite{Meyer-1988}.
    Classes can be extended with the \newsoda{\sodaextends} construct as follows

    \begin{center}
        \begin{tabular}{l}
            \sodaclass $A$ \\
            \sodaextends
            $B_{1}$
            $\ldots $
            $B_{n}$        \\
            \ \ $d_{1}$    \\
            \ \ $\ldots $  \\
            \ \ $d_{m}$    \\
            \sodaendclass
        \end{tabular}
    \end{center}
    where class $A$ extends classes $B_{1}, \ldots , B_{n}$, and $d_{1}, \ldots , d_{m}$ are constant or function definitions.

    It is also possible to define \emph{interfaces}, which are types that contain only declarations of constants and functions, without specifying what they contain.
    These declarations are in a block with the word \newsoda{\sodaabstract} as follows
    \begin{center}
        \begin{tabular}{l}
            \sodaclass $A$                \\
            \sodaabstract
            \ \ $f_{1}$ $\ldots $ $f_{n}$ \\
            \sodaendclass
        \end{tabular}
    \end{center}
    where each $f_{i}$ is a constant or function declaration.

    Each class has a default type constructor, which is named the same as the class with an extra underscore (`\_') as a suffix.
    The abstract elements in a class need to be given as parameters to instantiate a class.
    For example:

    \begin{lstlisting}[label={lst:example-of-pair}]
class Pair

  abstract
    fst : Int
    snd : Int

end
    \end{lstlisting}

    can be instantiated with $\srccode{Pair\_\ (1) (2)}$.

    Since we want to be able to apply design patterns and the language is statically typed and object-oriented, we need a way to refer to the instance itself, and the construct is \newsoda{\sodathis}.

    We allow type parameters for classes and functions to have polymorphism.
    To denote the type parameter, we use square brackets \newsoda{[]}.
    In class declarations, we specify that the parameter is of type $\srccode{Type}$.
    For example, a parameterized pair could be defined as
    \begin{lstlisting}[label={lst:example-of-parametric-pair}]
class Pair [A : Type] [B : Type]

  abstract
    fst : A
    snd : B

end
    \end{lstlisting}
    and then instantiated as $\srccode{Pair\_\ [Int] [Int] (1) (2)}$.

    In addition, it is possible to declare upper and lower type bounds.
    An upper type bound is denoted by the word \newsoda{\sodasubtype} or \newsoda{\sodalessthancolon}, as found in the literature, and the lower type bound is denoted by the word \newsoda{\sodasupertype} or \newsoda{\sodagreaterthancolon}.
    For example,
    \begin{center}
        \sodaclass $C$ \ \ [$A$ \sodasubtype $B$] \ \ \sodaextends $D$
    \end{center}
    indicates the declaration of class $C$ parameterized with type $A$, which is a subtype of $B$, and $C$ itself extends $D$.
    Subtyping in this language follows the Liskov Substitution Principle~\cite{Liskov-1994}.
    To organize the classes, we group them in \emph{packages},
    so that a package is just a collection of classes.
    We can declare that a class belongs to a package using the word \newsoda{\sodapackage}.
    In addition, the word \newsoda{\sodaimport} helps to bring classes from other packages.

    The language can be translated into other languages considering the so-called directives, which are specific to the target languages.
    The word \newsoda{\sodadirective} defines a piece of code that is considered only in specific translations and is ignored in others.

    \section{Discussion and Implementation}

    \subsection{Integration with Scala}

    To prototype the specification, we translate it into the Scala~\cite{Scala-2004} code.
    We use type checking and type inference provided by Scala.
    Each class (\sodaclass) is translated into a Scala \scalatrait, which is open for extension. Each constant definition is translated to a lazy value (\scalalazy \ \scalaval), and each function definition to a \scaladef, as well as each abstract constant or function declared with \sodaabstract.
    The default type constructor is a \scalacase \ \scalaclass that extends the original trait.
    Scala case classes provide constructors and extractors for pattern matching and they are not extensible.

    Each data type (\sodadatatype) is translated into Scala as a \scalasealed \ \scalatrait, having \scalacase \ \scalaobject for its constructors without parameters and \scalacase \ \scalaclass for its constructors with parameters.
    The \sodaif-\sodathen-\sodaelse and \sodamatch-\sodacase are very similar to what is provided in Scala.
    The supporting data types are the same as those provided by Scala, such as \scalaType{String}, \scalaType{Int}, \scalaType{Double}, \scalaType{Option}, and \scalaType{Seq}.

    The prototype can be run on the Java Virtual Machine (JVM), which is multiplatform and optimized for concurrent execution. It can, therefore, use JVM libraries so that, for example, a monitor can communicate with an AI agent interface.
    Although libraries can produce side effects, these can be easily controlled by the import commands, which define exactly the classes that are being used.
    On the one hand, a purely functional specification will not include any reference to those libraries.
    On the other hand, it is possible to connect an agent employing the corresponding JVM libraries.
    This dual use of the JVM brings the right amount of flexibility needed for practical use without losing control of critical parts.

    The translation to Scala comprises all the nuances of the language.
    Type parameters, in \Soda \ \scalaType{[A : Type]}, are also provided in Scala \scalaType{[A]}, as they are in Kotlin~\cite{Kotlin-2011} and Java \scalaType{<A>}, and Lean \scalaType{(A : Type)}.

    \subsection{Integration with Lean}

    Lean~\cite{Lean-2013} is a proof assistant and programming language based on the calculus of constructions with inductive types.
    Part of \Soda can be translated into Lean to prove the correctness of \Soda snippets.
    The types in Lean are not the same as those in Scala, and the JVM libraries cannot be used, but some core purely functional pieces of code in \Soda can be proven correct by using Lean.

    The class definitions in \Soda define three things that must be defined separately in Lean: a \emph{type}, a \emph{namespace}, and a \emph{constructor}.
    Every \Soda class defines a namespace in Lean \leannamespace. Some classes contain parametric internal values defined with \sodaabstract in \Soda. These classes are translated into Lean as \leanclass.
    They include a default constructor, which has the same name as the class, ending with an underscore, and the fields are all after the Lean \leanwhere .
    Lean already provides extractors needed for pattern matching.
    The default equality given in \Soda is achieved by deriving (\leanderiving) in Lean from \leanType{DecidableEq} (or from \leanType{BEq}).
    The definition of a data type can be directly translated into Lean by replacing the reserved word \sodadatatype with \leaninductive.

    As for Scala, the function definition in \Soda is translated to a \leandef at the beginning of the definition in Lean, \sodaif-\sodathen-\sodaelse in \Soda is identical in Lean, and a \sodamatch-\sodacase structure in \Soda is a \leanmatch-\leanwith-\leancasearrow structure in Lean.
    The structure
    \begin{center}
        \sodamatch $x$ \ \ \sodacase $p_{1}$ \sodacasearrow $e_{1}$ \ \ \ldots \ \ \sodacase $p_{n}$ \sodacasearrow $e_{n}$
    \end{center}
    is translated as
    \begin{center}
        \leanmatch $x$ \leanwith \ \ \leancasebar $p_{1}$ \leancasearrow $e_{1}$ \ \ \ldots \ \ \leancasebar $p_{n}$ \leancasearrow $e_{n}$
    \end{center}

    The case of package management (in \Soda \ \sodapackage and \sodaimport) and self-instances (in \Soda \ \sodathis) is not supported at the moment, and neither are the \sodasubtype and \sodasupertype type bounds.

    Some basic types in Lean are stricter than in \Soda or Scala, and there is not always a direct mapping from \Soda to Lean.
    However, it is possible to create a specific mapping with \sodadirective, which allows adding a mapping for a type, and including Lean theorems with their proofs.

    \subsection{Undefined States and Termination}

    \Soda does not use exceptions.
    This is because they correspond to an imperative feature, as they are used to interrupt the evaluation of a function and perform a jump (\emph{goto}) to the point where they are caught, assuming that they are caught at all.

    This also implies that designs in this language need to be careful, considering edge cases and properly managing them when building objects.
    If exceptions are thrown from JVM objects, they can be caught by underlying types in Scala (e.g. \scalaType{Try}), and then managed as objects.

    Aside from the possibility of integrating JVM libraries, the language itself is highly expressive.
    There is no limit to self-recursion, which brings advantages and disadvantages.
    We provide an annotation to force tail recursion (\sodatailrec) to avoid stack overflow.
    There are two prominent functions that can be easily defined in this language.
    The first function is \emph{range}, which generates the first natural numbers.
    The second function is \emph{fold}, which applies a cumulative operation on a sequence.
    Both functions are sufficient to define the most common operations on sequences.
    Since both functions above always terminate, they are a convenient substitute for full recursion, preventing possible infinite recursion.

    \subsection{Related Work}

    As mentioned above, the language aims to have a highly readable formal language for humans.
    In the design of the language, we consider the good properties of some programming languages.
    The main features we look for are:

    \begin{itemize}
        \item the specification is intended to be read and understood by a human (which is the most relevant point);
        \item everything defined is done only once, in one place (to avoid confusion due to partial definitions);
        \item objects are immutable (because one of the key properties is immutability);
        \item classes cannot be modified, but they can be extended (because of the Open-Closed Principle~\cite{Meyer-1988}).
    \end{itemize}

    One of the properties that we adopted was the \emph{functional notation}.
    For that, we evaluated three categories: the Lisp style, the ML (Meta Language)~\cite{ML-1973} style, and the Haskell style.
    In the Lisp category, we can mention Clojure~\cite{Clojure-2007}, which has a large community and can be integrated with the JVM.
    In the Haskell category, we can mention Haskell, which is the de-facto standard in the functional programming community.
    In the same category, we also have Agda~\cite{Agda-2007} and Idris~\cite{Idris-2007}, which can be used as proof assistants.
    In the ML category, we can mention OCaml~\cite{OCaml-1996}, which is a very efficient implementation of ML, Rocq (formerly Coq~\cite{Coq-1989}), which is a proof assistant with a very reduced core, and Lean~\cite{Lean-2013}, which is an efficient proof assistant.
    We wanted to avoid the excessive use of parentheses, as in the Lisp style, because it is less readable for non-experts.
    We wanted to avoid the definitions of partial functions, which is the standard notation in the Haskell style, in order to reduce undefined functions.
    We considered the ML style to be the most appropriate for the language, which induces definitions of total functions with a moderate number of parentheses.

    Another property we considered was \emph{readability}.
    For that we looked at Python~\cite{Python-1991,Peters-2004} and Prolog~\cite{Prolog-1972,SwiProlog-1987}.
    Python is a language that is popular among scientists without a computer science background and is directed at a broad range of ages.
    For example, some young students start with Scratch~\cite{Scratch-2007} and then transition to Python or JavaScript~\cite{JavaScript-1995}, while Scheme~\cite{Scheme-1975}, a minimalist dialect of Lisp, has been used to teach in university courses.

    Prolog is also widely accepted in the scientific community.
    For example, 2APL (A Practical Agent Programming Language)~\cite{Dastani-2008} is a programming language for multi-agent systems consisting of individual agents that may share and access external environments.
    2APL integrates the declarative and imperative style by connecting declarative beliefs and goals with events and plans.
    As later developments based on 2APL we can mention N-2APL (Norm-Aware Agent Programming Language), the 2OPL (Organisation Programming Language), and a framework for norm-aware multi-agent systems~\cite{Dybalova-2014} that integrates the programming languages.
    Although Prolog is a logic language and the interpreter itself operates in a different way, the pieces of code created in Prolog are similar to those in purely functional languages.

    Finally, we sought a good \emph{functional object-oriented integration}.
    This was provided by Scala, which has a thriving community of purely functional and object-oriented backgrounds.
    Scala also provides an advanced type inference system and can compile to JVM bytecode.

    \begin{table}[t]
        \caption{Properties of related languages.}
        \label{tab:properties-of-languages}

        \begin{center}
            \begin{tabular}{llcccccl}
                \toprule
                lang.    & version & A    & B    & C    & D    & E    & F         \\
                \midrule
                Agda     & 2.6.3   & \Yes & \Yes & \No  & \Yes & \No  & $10^{9}$  \\
                Clojure  & 1.11.1  & \Yes & \No  & \No  & \No  & \Yes & $10^{9}$  \\
                Rocq/Coq & 8.16.1  & \Yes & \Yes & \No  & \Yes & \No  & $10^{7}$  \\
                Haskell  & 8.8.4   & \Yes & \Yes & \No  & \Yes & \No  & $10^{8}$  \\
                Idris2   & 0.6.0   & \Yes & \Yes & \No  & \Yes & \No  & $10^{10}$ \\
                Lean     & 4.0.0   & \Yes & \Yes & \No  & \Yes & \No  & $10^{8}$  \\
                OCaml    & 4.14.1  & \Yes & \No  & \Yes & \Yes & \No  & $10^{9}$  \\
                Prolog   & 8.4.2   & \No  & \No  & \No  & \No  & \No  & $10^{8}$  \\
                Python   & 3.10.6  & \No  & \No  & \Yes & \No  & \No  & $10^{8}$  \\
                Scala    & 3.3.1   & \Yes & \No  & \Yes & \Yes & \Yes & $10^{10}$ \\
                \Soda    & 0.19.0  & \Yes & \Yes & \Yes & \Yes & \Yes & $10^{10}$ \\
                \bottomrule
            \end{tabular} \\
        \end{center}
        Columns:
        \textbf{A.} dominantly or only functional //
        \textbf{B.} purely functional in its core syntax //
        \textbf{C.} full object-oriented notation //
        \textbf{D.} statically typed //
        \textbf{E.} JVM integration //
        \textbf{F.} number of repetitions in 30 s
    \end{table}

    Table~\ref{tab:properties-of-languages} contains a summary of the properties we searched for.
    We tested the efficiency of the programming languages mentioned above on a computer with an Intel Core i5-8350U CPU (1.70 GHz) with 8 cores, running Linux 6.2.0 from the Ubuntu 22.04 LTS distribution.
    For all programming languages, we wrote a piece of code that was either a built-in cycle (as in Clojure, Prolog, and Python) or a recursion with tail recursion.
    We tried different powers of 10 and indicated the largest number of repetitions that fit within 30 seconds.
    This value is not meant to be a global benchmark to compare the languages or their implementations, but rather to show how \Soda is well-positioned in terms of efficiency.

    Employing formal proofs instead of only empirical evidence (such as unit tests) provides stronger reliability for multi-agent systems.
    This has been addressed in a development~\cite{Jensen-2021} where a verification framework for the GOAL agent programming language~\cite{Hindriks-2009} has been formalized in Isabelle/HOL~\cite{Nipkow-2002}, which is a proof assistant based on higher-order logic.
    We also became interested in Scallina~\cite{ElBakouny-2018-a,ElBakouny-2018-b}, which is a tool for translating from Rocq to Scala.
    As for languages for verification, we can also mention Dafny~\cite{Dafny-2009} and F*~\cite{FStar-2011}.

    \section{Conclusion}

    We present \Soda, a language used to model constraints in AI systems.
    Our main motivation for \Soda is to model complex requirements that need to be easily understood by humans.
    Furthermore, it can be used to model and prototype other types of constraints.
    In addition, \Soda can be efficiently prototyped in an optimized multiplatform state-of-the-art technology, like the JVM, and some pieces of code can be verified in Lean.

    We give an overview of a tool that helps to describe problems in a way that we consider more transparent and less error-prone.
    Although writing descriptions in this style could require more effort than with standard imperative languages, the effort to fully comprehend those descriptions is significantly smaller, thanks to the reduced number of constructs.
    This language is also conducive to writing better designs since each function explains a piece of code and serves as running documentation.
    In addition, the computer provides extra verification by checking and inferring the types.

    As future work, we can mention expanding the Lean translator to handle more nuances of \Soda.
    Also, since understandability is a key feature of \Soda, a case study in which stakeholders read descriptions can be used to corroborate the readability of the language.

    \subsubsection*{Acknowledgements}
    This work was partially supported by the Wallenberg AI, Autonomous Systems and Software Program (WASP) funded by the Knut and Alice Wallenberg Foundation.

    \bibliographystyle{splncs04}

    \bibliography{main}

\end{document}